\documentclass[aps,prb,superscriptaddress,twocolumn]{revtex4-2}

\usepackage{setspace}
\usepackage{lmodern}

\usepackage[latin9]{inputenc}
\usepackage{amsmath}
\usepackage{amssymb}
\usepackage{graphicx}
\usepackage{epstopdf}
\usepackage{color}
\usepackage{enumerate}
\usepackage{bm} 
\usepackage{physics}

\usepackage{array}
\usepackage[labelfont=bf]{caption}
\usepackage{multirow}
\usepackage{floatpag}
\usepackage[misc]{ifsym}
\usepackage{sistyle}
\usepackage{upgreek}

\begin{document}

\title{First-order quantum breakdown of superconductivity in amorphous superconductors}

\author{Thibault Charpentier}
\author{David Perconte}
\author{S\'{e}bastien L\'{e}ger}
\author{Kazi Rafsanjani Amin}
\author{Florent Blondelle}
\author{Fr\'{e}d\'{e}ric Gay}
\author{Olivier Buisson}
\affiliation{Univ. Grenoble Alpes, CNRS, Grenoble INP, Institut N\'{e}el, 38000 Grenoble, France}
\author{Lev Ioffe}
\affiliation{Google Research, Mountain View, CA, USA}
\author{Anton Khvalyuk}
\affiliation{Univ. Grenoble Alpes, CNRS, LPMMC, 38000 Grenoble, France}
\author{Igor Poboiko}
\affiliation{Karlsruhe Institute of Technology, Karlsruhe, Germany}
\author{Mikhail Feigel'man}
\affiliation{Univ. Grenoble Alpes, CNRS, LPMMC, 38000 Grenoble, France}
\affiliation{CENN Nanocenter, Lyublana 1000, Slovenia}
\author{Nicolas Roch}
\author{Benjamin Sac\'{e}p\'{e}}
\affiliation{Univ. Grenoble Alpes, CNRS, Grenoble INP, Institut N\'{e}el, 38000 Grenoble, France}

\begin{abstract}
\textbf{Continuous quantum phase transitions are widely assumed and frequently observed in various systems of quantum particles or spins. Their characteristic trait involves scaling laws governing a second-order, gradual suppression of the order parameter as the quantum critical point is approached. 
The localization of Cooper pairs in disordered superconductors and the resulting breakdown of superconductivity have long stood as a prototypical example. 
Here, we show a departure from this paradigm, showcasing that amorphous superconducting films of indium oxide undergo a distinctive, discontinuous first-order quantum phase transition tuned by disorder. 
Through systematic measurements of the plasmon spectrum in superconducting microwave resonators, we provide evidence for a marked jump of both the zero-temperature superfluid stiffness and the transition temperature at the critical disorder. 
This discontinuous transition sheds light on the previously overlooked role of repulsive interactions between Cooper pairs and the subsequent competition between superconductivity and insulating Cooper-pair glass.
Furthermore, our investigation shows that the critical temperature of the films no longer relates to the pairing amplitude but aligns with the superfluid stiffness, consistent with the pseudogap regime of preformed Cooper pairs. 
Our findings raise fundamental new questions into the role of disorder in quantum phase transitions and carry implications for superinductances in quantum circuits.}
\end{abstract}

\maketitle 

Superconductors undergo substantial changes in response to an increase in material disorder. Electron scattering, caused by disorder, increases resistivity and eventually leads to the breakdown of superconductivity due to Anderson localization and interactions~\cite{Sacepe20}. This breakdown, commonly referred to as the superconductor-to-insulator transition, has long been considered a prototypical continuous quantum phase transition~\cite{Sondhi97,Sachdev2011}, tunable by disorder, magnetic field or charge carrier density~\cite{Sondhi97,Goldman98,Gantmakher2010,Lin2015,Sacepe20}. 
	
The hallmark of the transition and of its quantum critical point is the gradual suppression of the superconducting order parameter following scaling laws with critical exponents~\cite{Fisher90,Sondhi97,Lin2015}. In disordered superconducting films, the central question has long been whether the amplitude (Cooper-pairing)~\cite{Finkelstein94} or the phase (macroscopic coherence)~\cite{Fisher90} of the superconducting order parameter is suppressed at the critical disorder, offering two different paths with distinct outcomes.

In recent years, a body of work on thin films of various materials has revealed a more subtle interplay between phase and amplitude suppression~\cite{Sacepe20}. Tunneling spectroscopy experiments have provided evidence of pairing amplitude persisting across the transition~\cite{Sacepe08,Sacepe11,Chand12,Sherman14}, indicative of the localization of Cooper pairs in the insulator, together with strong spatial fluctuations~\cite{Sacepe08,Sacepe11,Chand12,Kamlapure13,Carbillet20,Mandal20} and a pseudogap of preformed pairs \cite{Sacepe10,Mondal11,Sacepe11,Mondal13,Dubouchet18,Mandal20}. Concomitantly, this was accompanied by a substantial suppression of the superfluid (phase) stiffness~\cite{Mondal13,Pracht16,Singh18,Mandal20,Raychaudhuri21} upon approaching critical disorder, pointing to a prevailing role of phase fluctuations~\cite{Fisher90}. Yet, scaling laws, the hallmark of continuous QPTs, have never been demonstrated for the superfluid stiffness~\cite{Fisher89} in disorder-tuned transitions.

Here, we conducted a systematic study of the superfluid stiffness in one of the most disordered superconductors, amorphous indium oxide (a:InO) thin films, as we approach the breakdown of superconductivity by tuning disorder. Contrary to common expectations, we discovered that the superfluid stiffness does not exhibit scaling behavior with power-law suppression; instead, it shows a discontinuity at the critical disorder, signaling a first-order type quantum phase transition. Furthermore, we evidence that, at strong disorder, the superconducting transition temperature, $T_c$, is ruled by the superfluid stiffness in contrast with BCS superconductors, corroborating the pseudogap of preformed pairs. 

\begin{figure*}
 \centering
 \includegraphics[width=1\linewidth]{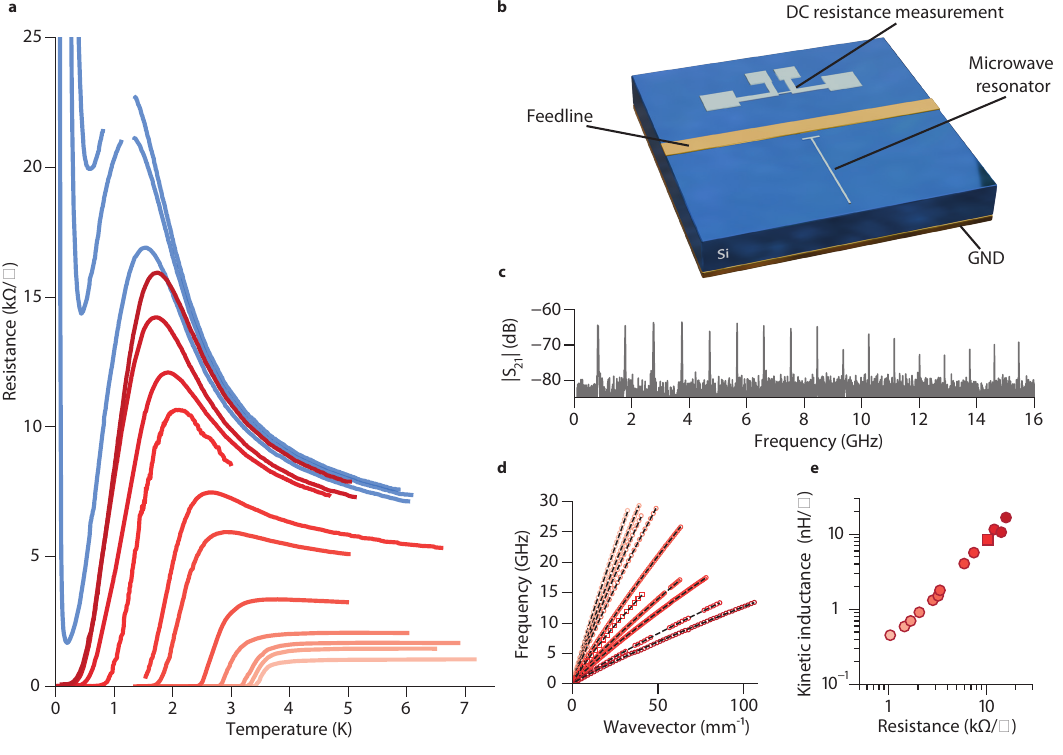}
 \caption{\textbf{Resistivity and microwave spectroscopy of indium oxide striplines.} \textbf{a.} Sheet resistance versus temperature for a series of amorphous indium oxide thin films. As sheet resistance increases, the critical temperature $T_c$ of the superconducting films (red curves) decreases up to the transition to insulator (blue curves). \textbf{b.} Schematics of the sample featuring a microstrip microwave resonator coupled to a feedline and a four-terminal device for resistance measurements. \textbf{c.} Typical two-tones measurement trace providing the frequencies of resonant modes up to 30 GHz, here truncated to 16 GHz for clarity. \textbf{d.} Dispersion relations of plasma modes extracted from two-tone measurements for films of increasing disorder. Dashed lines are fits following the theoretical dispersion relation of plasmons as described in Methods. The only fitting parameter is the film's kinetic inductance. \textbf{e.} Kinetic inductance per square of the films extracted from dispersion relations, as a function of sheet resistance maximum before the superconducting transition. Upon increase of disorder the kinetic inductance grows by nearly two orders of magnitude. Interestingly, it exhibits a power-law dependence with sheet resistance with exponent 1.4.}
 \label{fig1}
\end{figure*}  

\bigskip
\textbf{Superfluid stiffness}
\bigskip


At the core of this study is the systematic and accurate measurement of the superfluid stiffness, $\Theta $, together with the DC transport properties. The superfluid stiffness relates to the energy cost of twisting the superconducting phase $\varphi$, given by $E(\varphi) = \Theta \int d \mathbf{r} \frac{1}{2} |\nabla \varphi|^2$. Comparison of $\Theta$ with the other relevant energy scales of the superconducting state, such as the pairing gap and the superconducting transition temperature provides a direct assessment of the role of phase fluctuations. 
$\Theta$ smaller than $\Delta$ signals a phase-driven superconducting transition and the presence of preformed pairs~\cite{Emery95}. 
It is worth noting that the effective dimensionality depends on the physical phenomenon under consideration: in our samples, it is 3D for Cooper pairing, 2D for phase fluctuations and 1D for plasmons. See Methods for further discussion on dimensionality. \\

We designed superconducting microwave stripline resonators (see Fig. \ref{fig1}b) of a:InO enabling us to directly extract the kinetic inductance per square of the materials, $L_K$, through the superconducting plasmon dispersion~\cite{Mooij85,Camarota2001}. In the two-dimensional limit, the superfluid stiffness straightforwardly follows from
	\begin{equation}
\Theta =\left(\frac{\hbar}{2e}\right)^2 \frac{1}{L_K}
\label{eq1}
\end{equation}
($\hbar$ the reduced Planck constant, $e$ the electron charge). 

Figure \ref{fig1}a displays the superconducting transition of the sheet resistance of amorphous indium oxide thin films. Upon increasing disorder, characterized by the sheet resistance, the critical temperature continuously decreases (red curves), up to a critical value of resistance of $R_{\square}^c\simeq 16$ k$\Omega$ above which more resistive films show a drop of resistance and a re-entrant insulating behavior at the lowest $T$ (blue curves). In the same cooldown of each sample, we systematically performed two-tone microwave spectroscopy of the resonators' surface plasmons, by taking advantage of the intrinsic non-linearity of the superconductor (see Methods). The resulting plasmon modes shown in Fig. \ref{fig1}c are straightforwardly indexed in frequency $2\pi \, f_n = v k_n $ where $k_n=n\pi/L$ is the wavevector for mode $n$ and $v = 1 / \sqrt{lc_k}$ is the velocity of the mode ($L$ is the resonator's length, $l$ and $c_k$ are inductance and capacitance per unit of length, respectively) leading to the plasmon dispersions shown in Fig. \ref{fig1}d.

\begin{figure*}[ht!]
\centering
	\includegraphics[width=0.6\linewidth]{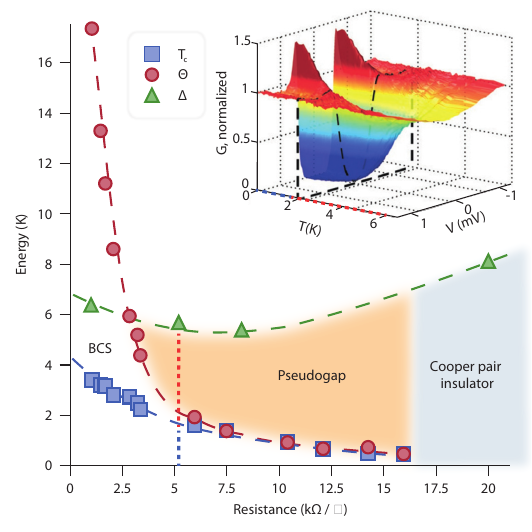}
	\caption{\textbf{Phase-driven superconducting transition.} Experimental phase diagram representing the critical temperature $T_c$, the single-particle tunneling gap $\Delta $ (from~\cite{Sacepe11,Sherman14,Sacepe15}) and the low temperature superfluid stiffness $\Theta $ as a function of sheet resistance. The tunneling data were obtained on virtually identical films \cite{Sacepe11,Sherman14,Sacepe15} allowing a consistent comparison with the current microwave samples. Upon increase of disorder, that is, resistance, the superconductor goes from a BCS regime with $\Theta > \Delta$ and $T_c \propto \Delta $ to a phase fluctuation regime characterized by $\Theta < \Delta$ and $T_c \approx \Theta$. In the latter, the superconducting transition is driven by the establishment of phase stiffness of preformed Cooper pairs that emerge as a pseudogap in the single particle density of states. Blue, green and red dashed lines are guide for the eyes. Inset: tunneling conductance measured on an a:InO film ($T_{\rm c}=1.7$ K) as a function of bias voltage across the tunneling junction and temperature. The black dashedline marks the tunneling spectrum at $T = T_c$. The pseudogap above $T_{\rm c}$ extends up to $\Delta \simeq 6$ K in excellent agreement with the phase diagram. Dotted lines below (blue) and above $T_c$ (red) on the inset are also reported in the figure at the corresponding disorder. Figure reproduced from Ref.~\cite{Sacepe11}.}
		\label{fig2}
\end{figure*}

Key to our analysis, the mode-dependent capacitance $c_k$ to the ground plane in our straight stripline geometry can be computed analytically (see Methods). Consequently, the sub-linear plasmon dispersions can be fitted with the inductance as a single adjustable parameter, yielding an accurate measurement of $L_K$. Consistent with expectations, we obtained a decrease of the plasmon velocity, that is, the slope of the dispersion in Fig. \ref{fig1}d, with increasing disorder. This reflects the increase of $L_K$ with sheet resistance shown in Fig. \ref{fig1}e reaching a maximum value of 17 nH$/\square$ at the transition to insulation. This value positions a:InO among the most inductive disordered superconductors, also known as superinductors~\cite{manucharyan2009fluxonium}, which exhibit wave impedance $Z=\sqrt{l/c_1}$ above the resistance quantum (see Extended Data Fig. \ref{extended_fig3}).

\bigskip
\textbf{Strong phase fluctuations of preformed pairs}
\bigskip

Translating the kinetic inductance into superfluid stiffness with Eq. (\ref{eq1}) enables us to construct the complete phase diagram of a:InO's superconducting quantum breakdown. In Figure \ref{fig2}, we present the three energy scales characterizing the superconducting state: the critical temperatures, $T_{\rm c}$'s, (see Methods for definition), the single-particle tunneling gaps, $\Delta $'s, from Refs.~\cite{Sacepe11,Sherman14,Sacepe15} and the superfluid stiffness, $\Theta $, as a function of sheet resistance. Notice the non-monotonic evolution of the tunneling gap that increases in the insulator as predicted by theory~\cite{Ma85,Ghosal01,Feigelman10b,Bouadim11}.

The dramatic effect of disorder on the superconducting order parameter is readily seen in the drop of $\Theta $. At low disorder, for $R_{\square}<3$ k$\Omega$, the superfluid stiffness is larger than the pairing energy, $\Theta > \Delta$, indicating that the phase is stiff and the superconducting transition is governed by the pairing of electrons at $T_{\rm c}$ according to BCS theory. However, for $R_{\square}>3$ k$\Omega$, the hierarchy of the two energy scales reverses: $\Theta < \Delta $. Strikingly, we also observe that $\Theta \simeq T_{\rm c}$ over a wide range of disorder, from approximately $7$ k$\Omega /\square$ --a value of the order of the resistance quantum for pairs $h/4e^2$-- up to $R_{\square}^c$. In this range of disorder where $\Theta < \Delta $, the superconducting transition is thus entirely governed by phase fluctuations. 

As earlier evidenced by tunneling spectroscopy and discussed in other contexts~\cite{Eagles69,Uemura89,Emery95}, the scenario at play is that of the preformation of Cooper-pairs at $T \sim \Delta $, followed at lower temperature by a wide regime of strong phase fluctuations also signaled by a pseudogap in the density-of-states~\cite{Sacepe11} (see inset in Fig. \ref{fig2}). Finally, at $T= T_{\rm c}\simeq \Theta $, the phase of the order parameter becomes stiff, establishing the quasi long-range order. 

Such a redefinition of $T_{\rm c}$ is a direct consequence of the very low superfluid density, the 2D effective dimensionality  (see Methods for dimensionality assessment) and the ensuing Berezinskii-Kosterlitz-Thouless transition~\cite{berezinskii72,kosterlitz73} observed in our data with a jump of superfluid density $\Theta (T_{\rm c}^{\rm BKT})=\frac{2}{\pi} T_{\rm c}^{\rm BKT}$ (see Extended data Fig.~\ref{extended_fig2}). The relation $\Theta (T=0) \simeq T_c$ naturally emerges in the 2D XY model and has been observed in some high-$T_c$ superconductors thin films~\cite{Lemberger11, Hetel07, Bozovic16}. More recently similar studies in NbN films~\cite{Chand12}, granular aluminum~\cite{Pracht16}, $\mathrm{LaAlO_3 / SrTiO_3}$ heterostructures~\cite{Singh18} or amorphous MoGe~\cite{Mandal20} showed that $\Theta$ approaches $T_c$ at strong disorder. Yet, a:InO stands out with an unprecedented large disorder range where $\Theta = T_{\rm c}$ (see Extended Data Fig.~\ref{extended_fig1}).

\begin{figure*}
           \includegraphics[width= \linewidth]{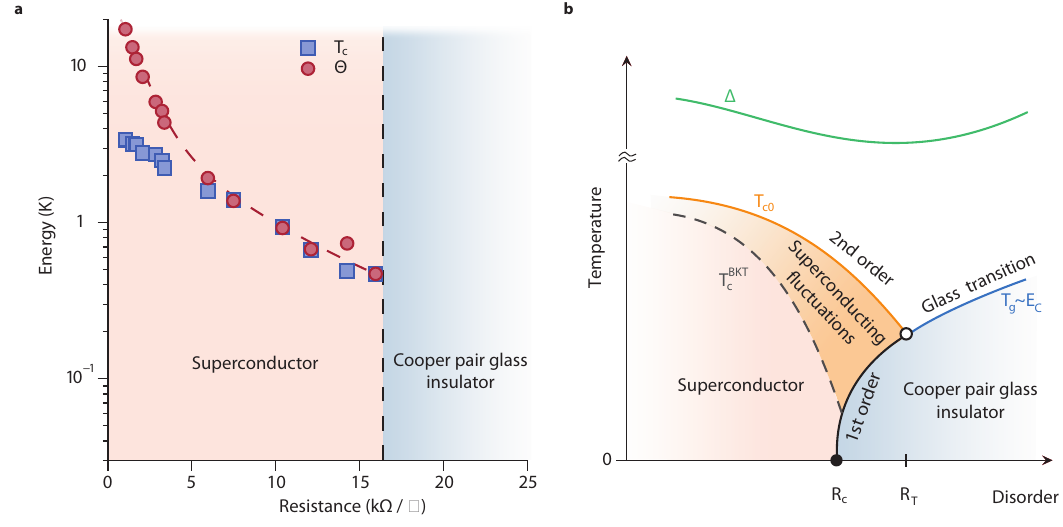}
\caption{\textbf{First-order quantum breakdown of superconductivity} 
\textbf{a.} Evolution of the low-temperature superfluid stiffness, $\Theta$, and critical temperature $T_{\rm c}$ with sheet resistance in semi-log scale. At the critical disorder indicated by the vertical black dashed line, both $\Theta$ and $T_{\rm c}$ remain finite, saturating at about $0.5$ K, without showing power-law suppression expected for quantum criticality in continuous quantum phase transitions. This abrupt, discontinuous suppression of superfluid density at the critical disorder indicates a first-order quantum phase transition. \textbf{b.} Phase diagram describing the competition between the superconducting phase and the Coulomb glass phase of localized Cooper pairs (see Methods for theoretical justification). The superconducting transition is characterized by the meanfield critical temperature $T_{c0}$ suppressed by disorder and interaction effects~\cite{Finkelstein94}, and the Berezinskii-Kosterlitz-Thouless transition temperature $T_c^{\text{BKT}}$~\cite{berezinskii72,kosterlitz73}, marking the onset of quasi-long range order. The single-particle gap $\Delta$ evolves non-monotonically across the transition, as observed experimentally (see Fig. \ref{fig2}) expected theoretically~\cite{Ma85,Ghosal01,Feigelman10b,Bouadim11}. $T_g$ is the glass transition temperature proportional to the Coulomb energy scale $E_{\rm c}$ (see Eq.~\eqref{condition1}), which evolves with disorder according to $E_C \approx 0.02 \,\delta_{\rm loc}$~\cite{FIC}, where $\delta_{\rm loc}$ denotes the mean level spacing in the localization volume. $R_c$ marks the critical disorder separating superconductor and Cooper pair glass, while $R_T$ is the resistance at the tri-critical point, where the three transition lines meet. The re-entrant insulating behavior emerges here due to the presence of a substantial region of temperature where thermal superconducting fluctuations~\cite{larkin2005} can decrease the resistance and mimic a superconducting transition, only to be aborted by the onset of the Cooper pair glass insulator.}
\label{fig3}
\end{figure*}  

\bigskip
\textbf{Superfluid stiffness discontinuity}
\bigskip

This understanding of the superfluid stiffness provides definitive and compelling evidence for the scenario of preformed Cooper pairs in a:InO~\cite{Sacepe11,Dubouchet18}. The question of their localization at the transition to insulator, which is central to this work, can now be addressed. We show in Fig. \ref{fig3} the same data for $\Theta$ and $T_{\rm c}$ in semi-log scale. While the continuous quantum phase transition paradigm would lead to a power-law suppression of $\Theta$ as one approaches the critical disorder, we instead observe a saturation of $\Theta $ at about $0.5$ K, which abruptly drops at the onset of the insulator. This drop translates in a disappearance of the resonance once the sample is insulating. Likewise, given that $\Theta \simeq T_{\rm c}$, the critical temperature exhibits a similar trend, also reaching approximately $0.5$ K before vanishing.
 
A discontinuity of $\Theta $ while the pairing amplitude remains finite is a clear indication of the breakdown of the macroscopic phase coherence, implying a suppression of the superconducting order parameter.
 The disorder-tuned transition to insulation in a:InO therefore undergoes a first-order type quantum phase transition. We conjecture that its origin lies in the overlooked Coulomb interaction between preformed (localized) Cooper-pairs. By incorporating long-range interactions between pairs, the insulator is expected to form a Coulomb glass of pairs, which, according to~\cite{MuellerIoffe,MuellerPankov}, possesses its own order parameter of the spin-glass type. The corresponding ground-state energy of the glassy state (per relevant volume) is given by the Coulomb energy $E_C$, which is proportional to the width of the Efros-Shklovsky Coulomb gap~\cite{ES}. In this scenario, the superconductor-insulator transition occurs between two ground states, each characterized by order parameters of distinct natures. The vanishing of the superconducting order parameter, coupled with the simultaneous appearance of the glassy order parameter, naturally manifests as an abrupt transition controlled by the competition between the free energies of these two very distinct phases of matter. In the $T=0$ limit, the condition for the energy of the superconducting state to be equal to the energy of the insulating Coulomb glass state can be expressed by the relation~\cite{PoboikoFeigelman2024}:
    \begin{equation}
     \Delta_c \approx E_C = \sqrt{\pi \nu_0} \frac{e^3}{\left( 4 \pi \varepsilon_0 \, \varepsilon\right)^{3/2}}
     \label{condition1}
 \end{equation}
 where $\Delta_c$ is the superconducting collective gap right before the transition to insulator, which is significantly different from the single-particle spectral gap~\cite{Dubouchet18}. Here, $\nu_0$ is the density of states at the Fermi level, and $\varepsilon$ is the macroscopic dielectric constant in the insulating state. Note that Eq.~\eqref{condition1} takes into account the fact that the elementary charge in our insulator is $2e$. 
It is important to highlight that the long-range Coulomb interaction driving the first-order transition is the interaction between bound electron pairs. The presence of a pseudogap in our strongly disordered films (see inset of Fig.~\ref{fig2} and references~\cite{Sacepe11,Dubouchet18}) provides experimental evidence for the existence of these pairs. Theoretically, Ref.~\cite{PoboikoFeigelman2024} explains why short-range electron repulsion~\cite{Finkelstein94} doesn't destroy the pairing itself. In short, this resilience is due to the fractal nature of electron wavefunctions in near-critical Anderson insulators.

The crucial consequence of Eq.~\eqref{condition1} is the existence of a maximum value of the kinetic inductance $L_K^{\text{max}}$ (or minimal value of superfluid stifness $\Theta^{\text{min}}$) achievable in a disordered thin-film superconductor. We can assess its order of magnitude by combining the semi-classical theory~\cite{Mattis58} relating $L_K$ to the superconducting gap and the normal-state resistance, with the phase transition condition Eq.~\eqref{condition1}, which leads to: 
 \begin{equation}
\Theta^{\text{min}} \approx \frac{g}{8} \, E_C 
\label{Lk-2}
\end{equation}
where $g=h/(e^2 R_{\square})$  is the dimensionless film conductance. With $\nu_0 = 2.4 \times 10^{46} \, \mathrm{J^{-1} m^{-3}}$ from
Ref.~\cite{Sacepe15} and $\varepsilon \approx 1000$ for a:InO~\cite{FIC,IF, Ebensperger21}, as well as $g \sim 1.5$, we estimate 
$\Theta^{\text{min}}$ to be approximately $0.4$ K, which closely aligns with the experimental value.

\bigskip
\textbf{Phase diagram}
\bigskip

The first-order nature of the transition is also reflected in the non-monotonic $T$-dependence of resistance in insulating samples near the critical disorder as shown in Fig. \ref{fig1}a. This re-entrant insulating behavior can be understood in terms of the temperature-dependent competition between the free energies of the superconducting and Coulomb glass phases. As depicted in the phase diagram presented in Fig. \ref{fig3}b, which is based on experimental observations and theoretically substantiated in the Methods section, one can observe that just above the critical resistance $R_c$, as the temperature decreases below the superconducting mean-field transition temperature $T_{c0}$, a region characterized by strong superconducting fluctuations --including the classical Berezinskii-Kosterlitz-Thouless mechanism-- emerges~\cite{larkin2005,Poboiko18}. Initially, these fluctuations lead to a reduction in resistance. However, as the temperature continues to decrease, the transition to the Cooper pair insulator glass occurs, thereby preventing the establishment of quasi-long-range order. The black solid line in Fig. \ref{fig3}b represents the first-order transition, originating from the quantum critical point (black dot) located at $R_c$, and extending to a tricritical point indicated by the open dot. Investigation into the physics surrounding the tricritical point would deserve further study.

A key implication of our work is that the properties of the disorder-tuned SIT depend on microscopic details of the material under consideration, and therefore are not universal. For instance, for a first-order transition to occur, two conditions must be met: the superconductor must present a pseudogap (like it is the case for NbN, TiN, grAl and some others), and the Coulomb gap amplitude must be comparable to the superfluid stiffness (as stated by Eq.~\eqref{Lk-2}). These two conditions strongly depend on parameters of the material such as dielectric constant and density of states.
A first-order transition is anticipated in disordered materials exhibiting an anomalously long electric screening length $l_{scr}$ comparable to the superconducting coherence length. In a:InO, calculations in Refs.~\cite{MuellerPankov,PoboikoFeigelman2024}, along with estimates for the dielectric constant $\varepsilon$ in the insulating state~\cite{FIC,IF}, suggest $l_{scr} \approx \xi_{loc}$, where the localization length $\xi_{loc}$ is approximately $4-5$ nm, close to that of the superconducting coherence length~\cite{Sacepe15}. Recent experiments in granular aluminum films also reported such a nm-scale screening length~\cite{Kristen23}. This material shares similar features with a:InO, such as a discontinuous drop in $T_c$ at the superconductor-insulator transition and non-monotonous re-entrant insulating state~\cite{LevyBertrand19}, suggesting a first-order breakdown of superconductivity, here too. Interestingly, $\mathrm{LaAlO_3/SrTiO_3}$ heterostructures~\cite{Singh18} reach 80 nH per square at the transition, i.e. $\Theta \sim 0.1$ K, which is consistent with a smaller Coulomb gap due to the large dielectric constant of SrTiO$_3$ known to be $\varepsilon \approx 10^4$.

\bigskip
\textbf{Discussion}
\bigskip

In a strongly disordered low-dimensional system, a first-order transition is not expected as disorder is known to smear the transition due to disorder-induced energy fluctuations and the pinning of domain walls that separate competing phases~\cite{ImryWortis,ImryMa,Vojta2013}.
However, in a three-dimensional (3D) system as ours, these effects do not necessarily  preclude the existence of a sharp first-order transition in the thermodynamic sense. Yet, observation of other distinct signatures of the first-order phase transition described above, such as glassy dynamics arising from the very slow motion of domain walls, would require the ability to continuously tune parameters such as magnetic field~\cite{Crane07} or carrier density via a gate voltage~\cite{Singh18}, which is beyond the scope of this study.

An important theoretical insight from our findings is the significant role of disorder in quantum phase transitions. Standard descriptions involve a mapping of QPTs in the D-dimensional system at T=0 onto the classical thermal transition in dimension D + 1 (where the "1" denotes the time dimension)~\cite{Sondhi97,Sachdev2011}. Such correspondence is possible in clean systems because world-lines in D+1 dimensional space-time are not correlated in the time domain. In contrast, strong, "frozen" disorder is by definition time-independent, leading to world-line correlations in D + 1 dimensional space-time. This makes disorder markedly more influential in altering the nature of a quantum phase transition compared to a clean system and thus may call to revisit our understanding of disorder-tuned QPTs.

Finally, the emergence of materials with very large kinetic inductance, resulting from suppressed superfluid stiffness, underscores their potential significance for both quantum circuits and sub-THz photon detectors~\cite{Doucot12, Grunhaupt19, Hazard19, Astafiev12, Siddiqi2021}. Superinductors capable of maintaining a quality factor of $\gtrsim 10^4$ (see Extended Data Figure \ref{extended_fig4} and Methods for a:InO quality factor), combined with their compact footprint, offer versatility across various applications, ranging from inductively shunted qubits to dissipative resonators and nonlinear parametric amplification, as well as highly sensitive photon detectors. Our findings establish an upper limit for the highest achievable kinetic inductance in disordered materials. Nevertheless, the precise role of disorder in bulk dissipation remains a critical aspect yet to be fully understood~\cite{Grunhaupt18}, promising to stimulate further investigation and reveal new fundamental insights.

\section*{Methods}

\subsection*{Samples}
Our samples are disordered thin films of amorphous indium oxide. The films with a thickness of $40$ nm are prepared by electron-beam evaporation of high-purity (99.999\%) $\mathrm{In_2O_3}$ onto a high resistivity silicon substrate while maintaining a controlled $\mathrm{O_2}$ partial pressure, enabling the tuning of disorder. Structures were patterned by electron-beam lithography on PMMA resist, followed by a development in IPA:$\mathrm{H_2O}$ at a temperature of 4 celsius. The sample backside is coated with a thick layer of gold to act as a ground plane for the microstrip resonator. The latter are mm-long and $1~\mathrm{\mu m}$-wide lines (see Table \ref{extended_table1} for exact sample geometries) and are accompanied with a co-evaporated Hall bar structure of same width (length $10~\mathrm{\mu m}$) allowing transport characterization of the films. 

\subsection*{Measurement setup}

The samples were placed in a copper sample holder shielded by a mu-metal and connected to the input and output microwave lines. DC lines are used for the resistance measurements. The a:InO resistance of the transport mesa structure (see Fig. \ref{fig1}b) was measured using standard lock-in amplifier technique and an ac current bias of $0.1-1$ nA. Microwave measurements were carried out in transmission with a vector network analyzer and a second microwave source for the two-tone measurements.

The technique employs a microwave resonator with a simple straight-strip line geometry, which can be probed at very low temperatures (20 mK), very low excitation powers (down to a few photons in average) and GHz frequencies (quantum regime $\hbar \omega \gg k_B T$). The cryostat in which the measurements took place is routinely used for the characterization of high-coherence superconducting circuits, providing a suitable filtered microwave environment. 
The originality of our experimental technique is the exploitation of the intrinsic non-linearities of the superconductor (Kerr effect) using two-tone measurements for the detection of resonance modes over a large frequency range, far beyond the frequency bandwidth allowed by standard microwave components.

\subsection*{Two-tone spectroscopy}

To obtain experimentally the dispersion relation of plasmons (shown in Fig.~1\textbf{d}) we perform a two-tone measurement, a technique that exploits the intrinsic non-linearity of current-phase relation in disordered superconductors. For a narrow superconducting wire of width $w$ and length $L$ at frequencies well below the gap $\omega \ll \Delta$, the latter can be described by the following Hamiltonian (see derivation in Appendix F of Ref.~\cite{Khvalyuk23}):
\begin{equation}
    H = \sum_{n}\hbar\omega_{n}^{\prime}\,a_{n}^{\dagger}a_{n}-\frac{\hbar}{2}\sum_{n,m}K_{nm}\,a_{n}^{\dagger}a_{n}\,a_{m}^{\dagger}a_{m},
    \label{eq:Kerr_Hamiltonian}
\end{equation}
where $a_{n}^{\dagger}$ and $a_{n}$ are bosonic creation and annihilation operators for the normal 1D plasmonic modes of the stripline, $\omega_{n}^{\prime} =\omega_{n}-\left(K_{nn}+\sum_{m}K_{nm}\right)/4$ is the angular frequency of plasmonic mode $n$ renormalized by the nonlinearity, and $K_{nm}$ is the Kerr coefficient.

A direct consequence of Eq.~\eqref{eq:Kerr_Hamiltonian} is the decrease of the observed plasmonic frequency $\omega_n$ when other modes $m \neq n$ are populated: $\omega_{n}\to\omega_{n}^{\prime}-\frac{1}{2}\sum_{m}K_{nm}N_{m}$ where $N_{m}=\langle a_{m}^{\dagger}a_{m}\rangle$ is the bosonic
occupation number of mode $m$. The two-tones spectroscopy technique exploits this effect to accurately resolve the plasmonic spectrum. Using a Vector Network Analyzer (VNA), the minimum of the transmission amplitude at a given mode $\omega_n$ is continuously monitored, while an external microwave source generates a signal varied from $100$ kHz to $20-30$ GHz. When the source frequency is far from a resonant mode of the stripline, the frequency of the transmission minimum is not shifted. As the source frequency approaches $\omega_m$, the frequency of mode $n$ decreases due to the Kerr effect. This translates into a sudden increase of the transmission amplitude at $\omega_n$ and results in an easily identifiable peak, as illustrated in Fig.~1\textbf{c}.

For a moderately disordered superconductor described by a dirty-limit semiclassical theory, the Kerr coefficients can be calculated analytically, yielding~\cite{Khvalyuk23}: 
\begin{equation}
    K_{nm}=3\gamma\left(1-\frac{1}{4}\delta_{nm}\right)\frac{\xi^{2}}{Lw}\,\frac{\hbar\omega_{n} \omega_{m}}{\Theta},
\label{eq:Kerr_coeff}
\end{equation}
where $\xi$ is the dirty limit superconducting coherence length, $\Theta$ is the superfluid stiffness, $\gamma = \frac{\pi}{4} + \frac{3}{4\pi}\approx 1.02$ is determined by the current-phase nonlinearity of a diffusive superconductor, and $\delta_{nm} = \begin{cases}0 \text{ if } n \neq m\\
1 \text{ if } n=m\end{cases}$. 

In this estimation we have assumed that the kinetic inductance fraction $\alpha = L_K / L_{\text{tot}} = 1$, i.e. the geometric inductance of the wire is negligible compared to the kinetic inductance, as is the case for disordered indium oxide. For a strongly disordered superconductor, the semiclassical description is rendered inapplicable both due to strong localization and due to the presence of pseudogap, hence the exact value of $K_{nm}$ is not known. However, one can still use Eq.~\eqref{eq:Kerr_coeff} to estimate the order of magnitude of the effect.

\subsection*{Extracting $L_K$ from the plasmon dispersion}

Surface plasmons in a thin, long and narrow ($d \ll w \ll L$) superconducting wire follow the sound-like dispersion relation $\omega_k = |k| / \sqrt{lc_k}$ where $l, c_k$ are inductance and capacitance per unit length respectively, and $k$ is the wavevector. The restoring force responsible for charge-density oscillations is the Coulomb interaction between distant charges, whose long-range character induces a weak $k$-dependence of the capacitance $c_k$~\cite{Mooij85}. In our particular geometry (displayed in Fig.1\textbf{b}) one also needs to account for screening effects of the silicon substrate (of thickness $h = 300~\mathrm{\mu m }$, with relative permittivity $\varepsilon = 11.9$) and the metallic ground plane underneath. This treatment can be found in Appendix F of~\cite{Khvalyuk23}. Here we show the resulting capacitance $c_k$ for two limiting cases:

\begin{equation}
\frac{1}{c_k} = 
\begin{cases}
\frac{1}{\pi \varepsilon_0 \left( 1 + \varepsilon \right)} \ln  \frac{8h}{we^{\delta(\varepsilon)}} & \text{if} \,  k \ll h^{-1}, \\
\frac{1}{\pi \varepsilon_0 \left( 1 + \varepsilon \right)} \ln  \frac{8}{kwe^{\gamma}} &\text{if} \,  w^{-1} \gg k \gg  h^{-1}, \label{eq:log_disp}
\end{cases}
\end{equation}
where $\gamma = 0.577...$ is the Euler-Mascheroni constant and
\begin{equation}
\delta(\varepsilon) = \frac{2 \varepsilon}{1 + \varepsilon} \sum_{j=1}^{\infty} \left( \frac{1 - \varepsilon}{1 + \varepsilon} \right)^{j-1} \ln \frac{1}{j}.
\end{equation}

At low $k \ll h^{-1}$ the screening is efficient and the dispersion relation is linear, $\omega(k) \propto k$, as first observed experimentally in Ref.~\cite{Camarota2001}. At larger $k \gg h^{-1}$ the dispersion curve bends down via a logarithmic correction, which was predicted in Ref.~\cite{Mooij85}. Eq.~\eqref{eq:log_disp} describes the plasmonic spectrum for $k \ll w$, i.e. in the region of 1D plasmons.  The open boundary conditions at each end of the stripline of length $L$ implies the quantization of the wavevector for resonant mode $n$ as $k_n = n\pi / L$ with $n = 1,\,  2\, ...$. 

The model above accounts accurately for the geometry of our samples, leaving the kinetic inductance per square $L_K = l w$ as the only fitting parameter. The reliability of extraction of $L_K$ has been confirmed by electromagnetic simulations in this geometry (see Supplementary Information I) and therefore offers high accuracy in determining the superfluid stiffness. This accuracy is further demonstrated by the excellent agreement between microwave and DC measurements.
Note that, contrary to most other experimental probes of the superfluid response (two-coil mutual inductance, scanning SQUID, microwave spectrometer...), we do not need any calibration of the experimental setup or background subtraction in order to extract the kinetic inductance.
We also note that the obtained $L_K$ are orders of magnitude larger than geometric inductances, hence confirming the assumption $\alpha = 1$.

\subsection*{Determination of critical temperature}

In a disordered superconductor with low superfluid stiffness, three distinct critical temperatures can be defined: first, the critical temperature predicted by BCS theory, denoted as $T_{\text{c0}}$; second, a temperature $T_{\text{c}} < T_{\text{c0}}$ that incorporates the effects of various superconducting fluctuations; and finally, the BKT transition temperature $T_{\text{BKT}} < T_{\text{c}}$, below which superconductivity breaks down due to the unbinding of vortex pairs. These last two temperatures depart upon increase of disorder, following~\cite{larkin2005}:
\begin{equation}
\frac{T_{\text{BKT}} - T_{\text{c}}}{T_{\text{c}}} \approx -4 \text{Gi},
\end{equation}
where $\text{Gi} = 7 \zeta(3) e^2 R_{\square} /(h \pi^3)$ is the Ginzburg-Levanyuk number and $R_{\square}$ is the normal-state sheet resistance measured above $T_c$.
The mean-field temperature is also suppressed, as 
\begin{equation}
T_{\text{c}} = T_{\text{c0}} \left( 1 - 2 \text{Gi} |\ln \text{Gi}|\right).
\end{equation}

We determine our critical temperatures ($T_c$) by identifying the temperature at which a linear extrapolation of the $R(T)$ curves intersects the $x$-axis (see Fig.~\ref{fig1}). While this method does not yield the exact determination of $T_{\text{BKT}}$, which typically requires a multi-parameter fit of the $R(T)$ curves (as demonstrated in Ref.~\cite{Weitzel23}), we are confident that our extracted values fall within the range $T_{\text{BKT}} \leq T_c^{\text{exp}} < T_{\text{c}}$. Moreover, measurement of the temperature-dependence of the superfluid stiffness shown in Extended data Fig.~\ref{extended_fig2} enabled us to accurately determine $T_{\text{BKT}}$ for a highly disordered sample (DP-res11). We concluded that our assessment of $T_c$ deviates from $T_{\text{BKT}}$ by less than 25\%. Hence, our methodology for estimating $T_c$ from transport measurements is sufficiently accurate for the discussions presented in the main text.

\subsection*{Effective dimensionality of a:InO films}

The thickness of our films is $d = 40$ nm, which is approximately an order of magnitude larger than the low-temperature superconducting coherence length $\xi(0)$. Consequently, the low-temperature behavior of our system should be described as effectively three-dimensional (bulk). Specifically, we applied the three-dimensional theory of the $T=0$ superconductor-Coulomb glass transition while deriving the condition in Eq.~\eqref{condition1}. However, the transition out of the superconducting state driven by temperature differs in terms of dimensionality. Given the very low superfluid stiffness of a:InO films in the pseudogap regime, long-range phase fluctuations emerge as the primary driving force of the transition, leading to a mechanism akin to vortex-antivortex de-pairing~\cite{Emery95}, reminiscent of the two-dimensional Berezinskii-Kosterlitz-Thouless transition~\cite{berezinskii72,kosterlitz73}.
Finally, our system is one-dimensional with respect to plasmons - collective excitations with wavelengths much longer than the 1-micron width $w$ of our superconducting stripes, see Fig.~\ref{fig1}d. This is the reason for these excitations to be ineffective in terms of thermodynamics: their density of states is too low in microscopic scale. In addition, the effective magnetic penetration depth is also much longer than $w$.\\
To summarize, one should consider our system to be effectively 1D, 2D or 3D depending on the mechanism at play. While long-wavelength plasmons responsible for microwave resonances and the Kerr effect are one-dimensional, the effect of thermal phase fluctuations near $T_c$ giving rise to the relation $T_c = \Theta$ is of a 2D nature, as expected from the BKT mechanism. Microscopic effects of disorder, local superconducting pairing and Coulomb repulsion occur on shorter scales than the film dimensions and are 3D in nature.


\subsection*{The phase diagram of a strongly disordered superconductor}

The determination of the position of the first-order Superconductor-Insulator Transition (SIT) at $T=0$, given by Eq.~\eqref{condition1} in the main text, stems from a comparison between the ground-state energy densities of the superconducting and insulating states, as shown in Ref.~\cite{PoboikoFeigelman2024}. The free energy density of the Coulomb glass state was calculated by solving the Parisi equations, and is expressed in terms of the key parameter $E_C$ defined by the right equality in Eq.~\eqref{condition1}. This expression for $E_C$ takes into account the charge $2e$ of a Cooper pair as well as the large dielectric constant $\varepsilon \simeq 1000$ of the underlying Anderson insulator of localized electrons~\cite{Ebensperger21,FIC,IF}. Note that the energy $E_C$ is proportional to (albeit differing by some numerical factor) the width of the Efros-Shklovsky Coulomb gap for this Cooper-pair insulator. While calculating the free energy of superconducting state in Ref.~\cite{PoboikoFeigelman2024}, we take into account the effect of the superconducting order parameter $\Delta$ upon the Coulomb screening energy: it leads to an increase of energy $\propto \nu_0 |\Delta| E_C$.
Such an effect is relevant in our problem while it is totally negligible in usual metallic superconductors. This is because the electric screening length in an Anderson insulator is comparable to both localization length and superconducting coherence length. A key assumption in our calculation was to treat the order parameter $\Delta$ as a constant throughout the entire system.  In fact, relatively close to the superconductor-insulator transition, the order parameter starts to fluctuate rather strongly from one point to another~\cite{Sacepe11}; for this reason, the left (approximate) equality in Eq.~\eqref{condition1} may contain an unknown factor of the order of unity.\\

The phase diagram presented in Fig.~\ref{fig3}b is the result of a delicate interplay between several phenomena: strong disorder of the superconducting phase, formation of the Coulomb glass in the insulating state, the physics of the resulting first order phase transition, and various manifestations of superconducting fluctuations in the insulating phase.

We start by discussing the phase diagram of a bulk superconductor (e.g., a film of very large thickness), where the broadening of the superconducting transition due to fluctuations is essentially absent. At low temperature, the phase transition happens between the superconductor and the Cooper pair glass insulator, thus being of first order. 

The transition line between the full and empty black dots in Fig.~\ref{fig3}b corresponds to the equality of the free energies of the two phases. The free energy of the Cooper pair glass, denoted as $\delta F_C(T)$, behaves as $F_C(T) - F_C(0) \sim - E_C (T/E_C)^4$, reflecting the quadratic shape of the soft Coulomb gap, $\nu(E) \propto E^2$~\cite{ES}. In contrast, the free energy of a strongly disordered superconductor, denoted as $\delta F_S(T)$, scales as $\sim - T_0 (T/T_0) ^{\beta + 1}$, where experimentally observed values are $T_0 \sim 10 \, \text{K}$ and $ \beta \sim 1.6$~\cite{Khvalyuk23}. This power-law suppression with temperature of the superfluid stiffness $\Theta$ in our samples, reported in Ref.~\cite{Khvalyuk23}, arises from the strong inhomogeneity of the superconducting state, allowing 3D small-scale low-energy excitations to contribute to the free energy~\cite{Khvalyuk23}.

For disorder slightly above the critical value (black dot in Fig.~\ref{fig3}b), the higher ground state energy of the superconducting state can be compensated at finite temperature by the higher entropy of this phase, resulting in $\delta F_S(T) < \delta F_C(T)$. Consequently, the first-order transition line initially trends towards higher temperatures with increasing disorder. In other words, the superconductor emerges as the high-temperature phase for this first-order phase transition. This contrasts with the classical BCS-like exponential dependence of the free energy, $\delta F_S(T) \propto -\exp(-\Delta / T)$, which would naturally lead to a conventional downward slope of the superconductor-insulator transition line.

Note that the above arguments rely on the quadratic energy dependence of the soft Coulomb gap $\nu(E) \propto E^2$, implicitly assuming the irrelevance of certain local two-level systems (TLS) typically found in glasses, which have a nearly constant density of states. While mean-field Coulomb glass theory~\cite{MuellerIoffe} does not incorporate such TLS, corrections to the mean-field approximation could result in their apparition. However, in our problem, deviations from this mean-field approximation are expected to be very small, on the order of $\sim E_C/E_F \ll 1$. Therefore, the apparent absence of these TLSs seems natural.

At sufficiently high disorder, on the other hand, the Coulomb glass turns into a trivial Cooper pair insulator via a glass transition~\cite{MuellerIoffe,MuellerPankov}, described by the transition line originating from the the tricritical point (empty dot) in Fig.~\ref{fig3}b. This inevitably implies the existence of yet another transition line (also starting from the empty dot on the phase diagram) separating the trivial insulator phase from the superconducting one. Here, the transition is of the second order, as observed experimentally~\cite{Sacepe11} and predicted by the existing theory of superconductor-insulator transition in the absence of Coulomb interaction~\cite{Feigelman10b}. This latter transition is described by the standard phenomenology, including the pronounced fluctuation effects in thin films. 

In particular, as one decreases the film thickness, the onset of superconductivity is pushed to lower temperatures  (dashed line in Fig.~\ref{fig3}b) by superconducting fluctuations and the Berezinskii-Kosterlitz-Thouless mechanism~\cite{larkin2005}, accompanied with a gradual drop in resistance as one approaches the superconducting phase. In parallel to that, the resistance of the Cooper pair glass insulator in a certain range of temperatures is decreased by the existence of superconducting puddles, inevitable consequence of the phase coexistence at the first-order phase transition. These two mechanisms underlie the intermediate drop of resistance of the insulating samples in Fig.~\ref{fig1}a, before this trend is sharply reversed by the phase transition to the Coulomb glass.

Importantly, however, the span of the first order transition line (between the full black and empty dots in Fig.~\ref{fig3}b) is expected to be rather short. Indeed, the overall magnitude of the effect of the aforementioned low-energy excitations in strongly disordered superconductor is small \cite{Khvalyuk23}, as evident by the fact that $T_0$ (featuring in the free energy decrease $\delta F_S$) tends to be 5 to 10 times larger than the superconducting transition temperature $T_{c0}$, and, via Eq.~\eqref{condition1}, $E_C$. The balance of the free energies then implies that the corresponding "anomalous" direction of the transition line is only expected in a narrow temperature window. Specifically, the temperature corresponding to the tricritical point (the empty dot in Fig.~\ref{fig3}b) is estimated as $T_T \lesssim E_C (E_C / T_0)^{\beta / (3 - \beta)} \ll E_C \sim T_{c0}$, with $\beta \sim 1.6$ and $T_0 / E_C \sim 7$ already rendering $T_T \lesssim 0.1 E_C$. Consequently, probing the vicinity of the transition point experimentally is challenging, as it is hard to continuously tune disorder.

\subsection*{Definition of disorder}

Quantifying the disorder of our samples presents a separate challenge due to the strong temperature dependence of the normal state resistance (see Fig.~\ref{fig1}a). This dependency prevents us from describing the level of disorder simply as resistivity at a certain temperature above the superconducting transition. Specifically, the sheet resistance of more disordered samples exhibits a local maximum, and the corresponding peak resistance, denoted as $R_{\square}$ in the main text, deviates from the room-temperature Drude resistance by a factor of up to $3-5$, resulting in an Ioffe-Regel number $k_F l_e \ll 1 $ (where $k_F$ is the Fermi wave vector and $l_e$ is the electron elastic mean free path).
However, $R_{\square}$ monotonously increases as the partial O$_2$ pressure at the stage of sample preparation is increased, providing a reliable measure of the final level of disorder in the sample. Moreover, for less disordered samples where the resistance above $T_c$ is nearly temperature-independent, this measure approaches the conventional one.
Finally, Fig.~\ref{fig1}e demonstrates that samples with different geometries but similar kinetic inductance per square exhibit similarly close $R_{\square}$ values, indicating that $R_{\square}$ is independent of sample geometry and thus accurately characterizes the bulk microscopic disorder.

\subsection*{Quantum phase slips rate estimation}

A discontinuity of $\Theta $ while the pairing amplitude remains finite echoes the superfluid jump at the quantum Berezinskii-Kosterlitz-Thouless transition in (1+1)D XY model. In this model, a superfluid jump is expected to occur at a critical wave impedance $Z_c=\sqrt{l_c/c_1}=\frac{1}{3}\frac{h}{4e^2}$ \cite{Giamarchi88}, where $l_c$ is the critical inductance (superfluid stiffness) at the jump. This jump marks the transition from a superfluid state to a Bose-glass phase, characterized by the proliferation of quantum phase slips~\cite{Kuzmin19}. In our resonators, the surface plasmons are indeed 1D electromagnetic modes. 

The quantum phase slip rate in superconducting wires can be calculated knowing the superconducting gap $\Delta$, wire dimensions and normal state resistance per square $R_{\square}$~\cite{Arutyunov08}. In strongly disordered superconductors with a pseudogap of preformed pairs such expression cannot hold and one must instead turn to the framework described in Ref.~\cite{Feigelman10b}, leading to the estimate of phase-slip amplitude~\cite{Astafiev12}:
\begin{equation}
h \gamma_{\text{QPS}} \approx \Theta \sqrt{\frac{L}{w}} \exp \left( -\eta w \sqrt{\Theta \nu d} \right)
\end{equation}
where $\eta \sim 1$ is a dimensionless constant, $\Theta$ is the 2D superfluid stiffness, $d$ the film thickness and $\nu$ is the single-particle density of states. Using $\nu = 2.4 \times 10^{46} \, \mathrm{J^{-1} m^{-3}}$ for indium oxide, $d = 40$ nm and width $w=1~\mathrm{\mu m}$ gives the estimation $h \gamma_{\text{QPS}} \approx  10^{-34}~\text{K}$ for our most disordered sample (having $\Theta = 0.5$ K). The reason for such a low phase-slip rate is the large cross section of our films. In comparison, a:InO nanowires in Ref.~\cite{Astafiev12} had width $w = 40$ nm and a phase-slip amplitude $h \gamma_{\text{QPS}} \approx 0.2$ K.

Consequently, the quantum phase slip rate $h \gamma_{\text{QPS}}$ is completely negligible due to the large width ($1\,\mu$m) of our resonators, thus excluding this scenario. Furthermore, the wave impedance of all our resonators are well above $Z_c$, as shown in Extended Data Fig. \ref{extended_fig3}, reaching up to $Z\approx 3\frac{h}{4e^2}$ at the critical disorder.

\subsection*{Dissipation in a:InO resonators}

Dissipation in our resonators manifests through the value of the internal quality factor, which is extracted from fitting the microwave transmission spectrum (Supplementary Information and Fig.~S2).

Extended data figures \ref{extended_fig4}a and b display the typical evolution of $Q_i$ as a function of photon numbers and temperature, respectively. As a function of photon number, $Q_i$ increases continuously. As a function of temperature, $Q_i$ is non-monotonic: it initially decreases when warming up, then increases up to $T \approx 0.4$ K and decreases at higher $T$. Such dissipation behaviors are commonly accounted for by the presence of a bath of two-level systems (TLS) that saturate upon increasing photon number or temperature~\cite{Muller19}, with the decrease at higher $T$ being attributed to thermally-activated quasiparticle dissipation~\cite{Mattis58}. Overall, for most a:InO resonators, we obtained $Q_i \approx 10^4$ at low photon number and low temperature.

Contrary to clean superconductors, dissipation in our a:InO resonators is not limited by surface TLS's. This is evidenced by the fact that resonators measured in a 3D cavity, for which the surface participation ratio is reduced by one order of magnitude, exhibit the same quality factor (see Extended data Fig.~\ref{extended_fig4}). Furthermore, to exclude the possibility of surface oxide contribution, we studied similar stripline resonators capped with a thin oxidized aluminum layer, which showed no change in $Q_i$ (see Extended data Fig.~\ref{extended_fig4}a). This leads to the conclusion that dissipation in a:InO is predominantly bulk-related and therefore associated to disorder, in accordance with other inductive materials~\cite{Grunhaupt18, Kristen23, Amin22}.

\bigskip

\section*{Acknowledgments}

We thank D. Basko, L. Benfatto, J. Delahaye, T. Grenet, V. Kravtsov, M. Muller, M. Scheffler, and C. Strunk for valuable discussion. 
We thank E. Eyraud for assistance with cryogenics and J. P. Martinez for preliminary measurements. 
T.C. and B.S. acknowledge funding from the ANR Project No. ANR-19-CE30-0014-CP-Insulators. A.V.K. is grateful for the support by Laboratoire d'excellence LANEF in Grenoble (ANR-10-LABX-51-01). B.S. has received funding from the European Union's Horizon 2020 research and innovation program under the ERC grant SUPERGRAPH No. 866365. N.R. has received funding from the European Union's Horizon 2020 research and innovation program under the ERC grant SuperProtected No. 101001310. N.R., B.S. and D.P. acknowledge funding from the ANR agency under the 'France 2030' plan, with Reference No. ANR-22-PETQ-0003. I.P. acknowledges support by the Deutsche Forschungsgemeinschaft (DFG) via the grant MI 658/14-1.

\section*{Competing Interests} The authors declare that they have no competing interests.

\bibliography{Biblio}


\clearpage

\setcounter{figure}{0}
\renewcommand{\figurename}{Extended Data Fig.}
\renewcommand{\tablename}{Extended Data Table}


\begin{table*}[hb!]
\caption{Summary of experimental data discussed in the text. $w$ and $L$ are resonator width and length respectively, $T_c$ is the critical temperature (as discussed in Methods) and $R_{\square}$ is the maximum of sheet normal-state resistance measured just above $T_c$. $L_K$ is obtained via fitting of plasmon dispersion relation (see Methods), allowing to estimate the superfluid stiffness $\Theta = (\hbar / (2 e))^2 /L_K$. Note that the sample (DP-Res11, represented by squares in Fig.~\ref{fig1} \textbf{d} and \textbf{e}) has different dimensions than the others, and thus has a dispersion relation that does not follow the disorder-evolution of the other samples, as displayed in Fig.~\ref{fig1} \textbf{d}. The relation between resistance and kinetic inductance shown in Fig.~\ref{fig1} \textbf{e} however is consistent with the other samples.}
\centering
\begin{tabular}{|c||c|c|c|c|c|c|}
\hline
Sample  & $w~\mathrm{(\mu m)}$ & $L~\mathrm{(m m)}$ & $T_c~\mathrm{(K)}$ & $R_{\square}~\mathrm{(k\Omega / \square)}$ & $L_K~\mathrm{(nH/ \square)}$ & $\Theta~\mathrm{(K)}$  \\
\hline
TC002-2 & 1 & 3.505 & 3.4 & 1.04 & 0.452 & 17.35 \\  
TC002-1 & 1 & 3.505 &3.2 & 1.456 & 0.59 & 13.29 \\  
TC014 & 1 & 2.5 &3.16 & 1.683  & 0.70 & 11.2 \\ 
TC003       & 1 & 3.5 &2.8 & 2.06  & 0.91 & 8.6 \\  
TC040 (3D)       & 1 & 2 &2.74 & 2.84 & 1.32 & 5.94 \\ 
TC007-3 & 1  & 3.505 &2.5 & 3.22 & 1.51 & 5.2 \\  
TC001       & 1  & 3.505 & 2.24 & 3.36 & 1.79 & 4.38 \\  
TC007-2 & 1 & 3.505 &1.6 & 5.95 & 4.06 & 1.93 \\  
TC007-1 & 1 & 3.505 &1.4 & 7.47 & 5.68 & 1.38 \\  
DP-res11 & 5 & 1.39 & 0.94 & 10.65 & 8.48 & 0.92 \\
TC016-8 & 1 & 1.718 &0.67 & 12.1 & 11.6 & 0.68 \\  
TC017       & 1& 3.505 &0.49 & 14.25 & 10.65 & 0.73 \\  
TC016-6 & 1& 1.718 &0.47 & 15.95 & 16.68 & 0.47 \\
\hline
\end{tabular}
\label{extended_table1}
\end{table*}

 \begin{figure}[ht!]
            \centering
 \includegraphics[width=\linewidth]{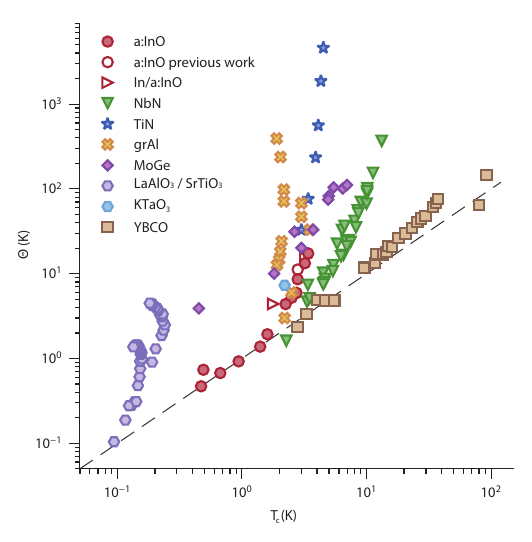}
 \caption{\textbf{Evolution of superfluid stiffness with critical temperature for various superconductors.} Upon variation of disorder, carrier density or doping, superconductors display a decrease of both superfluid stiffness and critical temperatures. For very low superfluid densities these two quantities become of the same order, as evidenced by the dashed line representing the equality $\Theta = T_c$. a:InO lies on this line in a large disorder range. Data from~\cite{Chand12, Yong13, Weitzel23} (NbN), \cite{Pracht16, LevyBertrand19} (grAl), \cite{Bert12, Singh18} ($\mathrm{LaAlO_3 / SrTiO_3}$), \cite{Mallik22} ($\mathrm{KTaO_3}$), \cite{Misra13, Mandal20, Turneaure01} (MoGe), \cite{Amin22} (TiN), \cite{Basov95, Hetel07} (YBCO). Previous works on a:InO \cite{Dupre17} and In/a:InO composites \cite{Fiory83, Turneaure01} are also added.}
 \label{extended_fig1}

  \end{figure}      
 \begin{figure}[ht!]
            \centering
 \includegraphics[bb=1cm 0cm 8.5cm 7cm,width=1\linewidth]{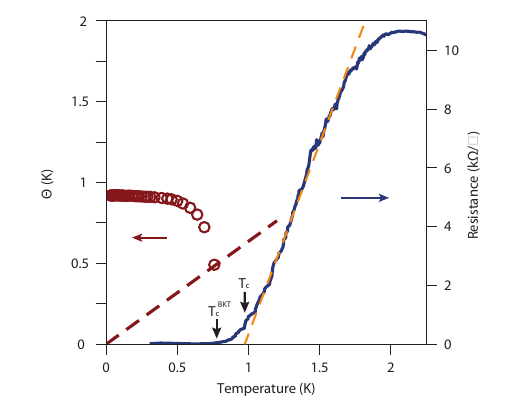}
 \caption{\textbf{Berezinskii-Kosterlitz-Thouless transition of sample DP-res11.} Left axis displays the superfluid stiffness versus temperature obtained from the $T$-dependence of the frequency shift of plasma modes via $\Theta(T) = \Theta(0) (f(T) / f(0))^2$. Dashed red line represents the Berezinskii-Kosterlitz-Thouless universal critical line $\Theta(T) = \frac{2}{\pi} T$. Both curves cross exactly at the vortex unbinding temperature $T_{\text{BKT}} = 0.75$ K. Right-axis shows the corresponding superconducting transition in the sheet resistance versus temperature. The dashed orange line illustrates our definition of $T_c$ as a linear extrapolation of the resistance curve, giving $T_c = 0.94$ K.}
 \label{extended_fig2}
  \end{figure}


\begin{figure}[ht!]
     \centering
           \includegraphics[bb=1cm 0cm 8.5cm 7cm,width= 1.\linewidth]{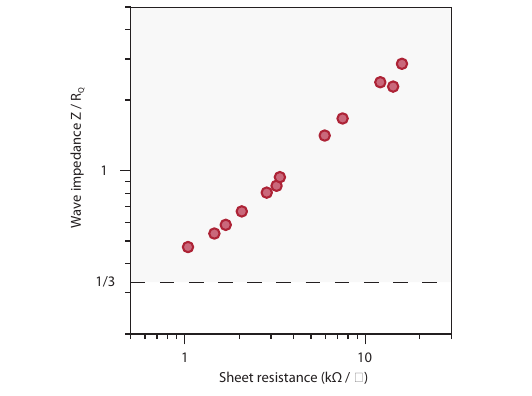}
\caption{\textbf{Superinductance.} Wave impedance $Z = \sqrt{l/c_1}$ versus sheet resistance. The impedance is normalized by $h/4e^2$. All data points lie well above the line $Z / (h/4e^2) = 1/3$, and some of them even achieve $Z > h/4e^2$. This classifies them as superinductances.}
\label{extended_fig3}
\end{figure}   

 \begin{figure*}[ht!]
            \centering
 \includegraphics[width=1\linewidth]{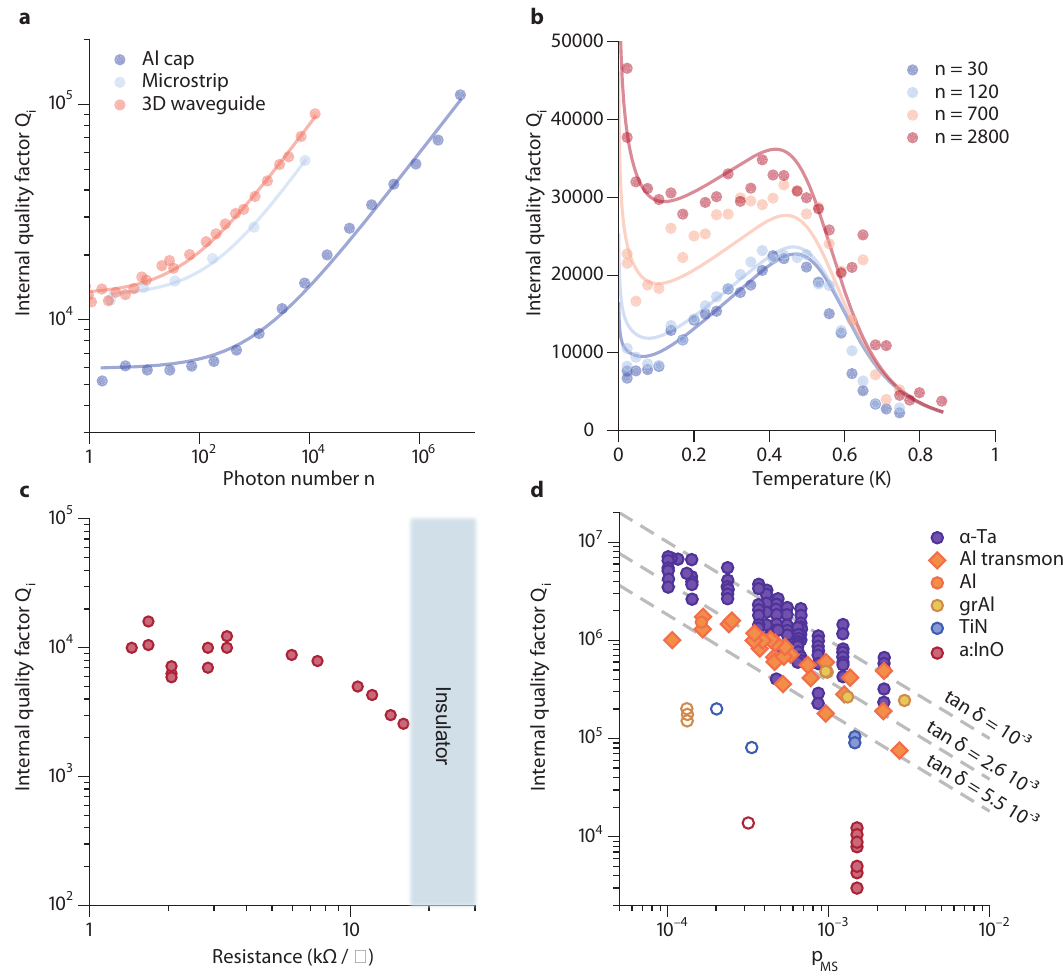}
 \caption{\textbf{Microwave dissipation in a:InO superconducting resonators.} \textbf{a} Evolution of a:InO resonators quality factors upon increasing power for three different sample environments, with varying sensitivity to surface dielectric loss. We studied resonators in the microstrip geometry (see Fig.~\ref{fig1}) or embedded in 3D aluminum waveguides (See Fig. S3) with reduced surface loss participation ratio. A third type of sample is capped in-situ after a:InO deposition by a thin aluminum layer to replace the surface dissipation of a:InO by thin aluminum oxide. \textbf{b} Temperature evolution of an a:InO resonator (sample TC040) in 3D waveguide showing non-monotonous behavior. Solid lines in panels \textbf{a} and \textbf{b} are fits following a TLS model (see SI). \textbf{c} Evolution of the low-power, low-temperature quality factor with sheet resistance, for the samples reported in Extended Table~\ref{extended_table1}. Near the transition to insulator the quality factor remains $>2 \times 10^{3}$. \textbf{d} Evolution of low-power and low-temperature quality factor with metal-substrate participation ratio (see SI). Full symbols correspond to resonators and transmon qubits in a 2D geometry (Ta~\cite{Crowley23}, Al and grAl~\cite{Grunhaupt18}, TiN~\cite{Amin22}, Al transmons~\cite{Wang15}, a:InO), empty symbols show resonators measured in a 3D waveguide (grAl~\cite{Grunhaupt18}, TiN~\cite{Amin22} and a:InO sample TC040). Dashed lines show the expected scaling for dielectric loss $Q_i = \left[p_{\text{MS}} \tan \delta \right]^{-1}$ for three values of $\tan \delta$.}
 \label{extended_fig4}
  \end{figure*}

\clearpage
\setcounter{figure}{0}
\renewcommand{\figurename}{SI Data Fig.}
\renewcommand{\tablename}{SI Data Table}

\bigskip
\section*{Supplementary Information}
\bigskip

\section{Electromagnetic simulation of plasmon dispersion relations}

\begin{figure}[ht!]
           \includegraphics[bb=1cm 0cm 8.5cm 7cm,width= 1\linewidth]{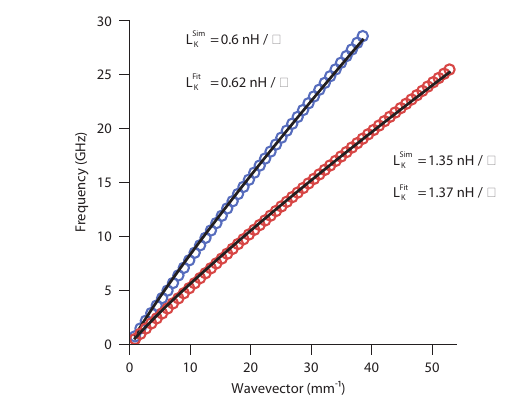}
\caption{\textbf{Simulated plasmon dispersion and fit.} SONNET simulations of two $3505~\mathrm{\mu m}$-long and $1~\mathrm{\mu m}$-wide microwave resonators on a $300~\mathrm{\mu m}$-thick substrate ($\varepsilon = 11.9$) with two different surface inductances $L_K^{\text{Sim}} = 0.6~\mathrm{nH / \square}$ (blue points) and $L_K^{\text{Sim}} = 1.35~\mathrm{nH / \square}$ (red points). Solid black lines are fit using our model, with only fitting parameter $L_K^{\text{Fit}}$. Our extraction of $L_K$ agrees very well with the simulated value, within 3 \%. }
\label{supp_mat_fig4}
\end{figure}

We cross-checked the validity of our model for the plasmon dispersion relation by simulating the response of a resonator in a geometry reproducing the experimental conditions. In SONNET software one can set a surface inductance $L_K^{\text{Sim}}$, emulating the material's sheet kinetic inductance $L_K$. Fig.~\ref{supp_mat_fig4} shows the resulting dispersion relations for two values of $L_K^{\text{Sim}}$, with frequencies up to 30 GHz.

Just like the experimental data, the simulated plasmon resonances exhibit a sub-linear dispersion relation upon increase of wavevector $k = n\pi / L$. Using our plasmon model we obtain inductances $L_K^{\text{Fit}}$ in excellent agreement with the reference values set in the simulation, with an error smaller than 3 \%.

\section{Supplementary two-tones spectroscopy data}

 \begin{figure}[ht!]
            \includegraphics[width= 1\linewidth]{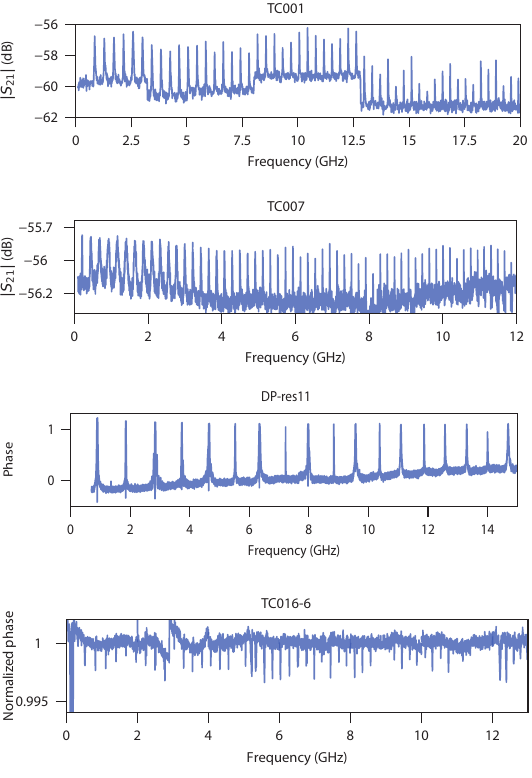}
\caption{\textbf{Two-tones spectroscopy spectrum for four samples of increasing disorders.} From top to bottom: resonance modes obtained via the amplitude or phase response of two-tones spectroscopy for increasingly disordered samples TC001, TC007, DP-res11 and TC016-6. Data for TC016-6 is filtered and a background is removed.}
 \label{supp_mat_fig5}
 \end{figure}



Figure~\ref{supp_mat_fig5} shows additional two-tone spectroscopy data for four other samples with increased disorder (TC001, TC007, DP-res11, and TC016-6), complementing Figure 1c which displayed data for sample TC014. The resistances and geometries of these samples are provided in Extended Data Table 1. Either the phase or the amplitude of the transmission coefficient $S_{21}$ is represented, depending on which parameter exhibited the best signal-to-noise ratio. Whether a dip or a peak is observed in the transmission depends on the frequency of the probe tone (i.e., whether it sits exactly at the resonance frequency or is slightly detuned).

For a given geometry, the plasmon spectrum is given by $2\pi f_n = vk_n$ where $k_n=n\pi/L$ is the wavevector for mode $n$ and $v = 1 / \sqrt{lc_k}$. Therefore, at low $k$, the mode spacing is roughly constant $\sim v \pi / L$, which is close to the fundamental mode frequency $2\pi f_1 = v \pi / L$. At a given wire width, the fundamental frequency and mode spacing are controlled by two parameters: the kinetic inductance per unit length $l$ and the resonator length $L$. As disorder increases, $l$ increases, which reduces the frequency. However, our resonators do not all have the same length (see Extended Data Table 1). In particular, our most strongly disordered resonator just before the SIT (sample TC016-6) is shorter than most others, which allows for a mode spacing large enough to resolve multiple resonances.


\section{Extraction of quality factors}

\begin{figure}[ht!]
     \centering
           \includegraphics[width= 0.8\linewidth]{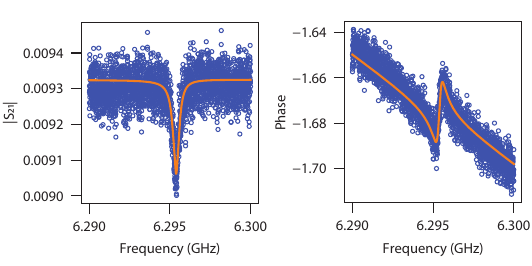}
\caption{\textbf{Example of resonance at single photon power.} Sample TC040 was measured in a 3D waveguide. We show both amplitude and phase of the complex scattering parameter $S_{21}$. Solid lines are a fit to the data from which we extract $f_r = 6.295$ GHz, $Q_i = 1.48 \times 10^4$, $Q_c = 5.13 \times 10^5$. }
\label{supp_mat_fig3}
\end{figure}

We utilize the method and Python package provided in \cite{Probst2015} to fit the complex transmission scattering parameter $S_{21}$ near the resonance frequency $f_r$. This approach enables an efficient extraction of resonator parameters from noisy experimental data, as demonstrated in Fig.~\ref{supp_mat_fig3}. For most of our samples, the coupling quality factor $Q_c$ significantly exceeds $Q_i$, ensuring that the total loss is primarily dictated by the indium oxide resonator rather than its environment. The formula for $S_{21}(f)$ is given by:
 \begin{equation}
 S_{21}(f) = \underbrace{a e^{j \eta} e^{-2 \pi j f \tau}}_{\text{environment}} \underbrace{\left[ 1 - \frac{(Q_l / |Q_c|) e^{j\phi}}{1 + 2 j Q_l (f/f_r - 1)} \right]}_{\text{ideal resonator}}
 \label{eq:formule_S21}
 \end{equation}
Here, $Q_c$ represents the coupling quality factor, $Q_l = \left[ Q_i^{-1} + Q_c^{-1}\right]^{-1}$ denotes the loaded quality factor, and the phase $\phi$ accounts for impedance mismatch along the line. Apart from these parameters describing an ideal coupled and lossy resonator, additional losses due to the environment are characterized by defining an amplitude $a$, phase shift $\eta$, and electric delay $\tau$, which consider the finite wave velocity in a cable of finite length.

The average photon number circulating in the resonator at resonance $f_r = \omega_r / (2 \pi)$ can be estimated by $\displaystyle n \approx 4 P_{\text{in}} Q_{\text{l}}^2 / (\hbar \omega_r^2 Q_c)$ where $P_{\text{in}}$ is the on-chip microwave power and the loaded quality factor is $Q_l = \left[ Q_i^{-1} + Q_c^{-1}\right]^{-1}$.

\section{Model for quality factors as function of temperature and power}

The model used to fit the quality factors shown in Extended Fig. 4a and b consists of three dissipation channels added in parallel:
\begin{equation}
\frac{1}{Q_i} = \frac{1}{Q_\text{TLS}(n,T)} + \frac{1}{Q_\text{QP}(T)} + \frac{1}{Q_\text{other}}
\label{eq:TLS_1}
\end{equation}
where $n$ is the average photon number circulating in the resonator, and $T$ is the temperature. The first term of Eq.~\eqref{eq:TLS_1} explicitly depends on both $n$ and $T$, the second one on temperature only, and the last one does not depend on either and can be seen as a saturation value of unknown origin.

 \subsection{Thermal quasiparticles}
 At relatively large temperatures one expects thermal (equilibrium) quasiparticles to be a significant dissipation channel. The latter is described by the Mattis-Bardeen theory~\cite{Mattis58}, and can be written
 \begin{equation}
 Q_{\text{QP}} = A_{\text{QP}} \frac{e^{\Delta / k_B T}}{\sinh \left( \frac{\hbar \omega}{2k_BT} \right) K_0 \left( \frac{\hbar \omega}{2k_BT} \right)}
 \label{eq:Q_QP}
 \end{equation}
 where $A_{\text{QP}} \approx \pi / (4 \alpha)$ is inversely proportional to the kinetic inductance fraction $\alpha = L_k / L_{\text{tot}}$, $\omega / (2 \pi)$ is the resonance frequency, $K_0$ is the zeroth-order modified Bessel function of the second kind and $\Delta$ is the superconducting gap.\\

 Eq.~(\ref{eq:Q_QP}) is in good agreement with the exponential decay of $Q_i(T)$ for $T > 500$ mK, as seen in Extended Data Fig.4b.
 At lowest temperatures the density of equilibrium quasiparticles vanishes: upon cooling down $Q_{\text{QP}}$ should increase continuously. However it is well known experimentally that the low-$T$ quality factor of superconducting resonators always remains finite, and even increases with temperature or microwave power. We also observe a similar trend for a:InO, as depicted in Extended Data Fig. 4a and b.\\

 \subsection{Two-Level Systems}
 One usually considers an additional loss mechanism due to coupling to an ensemble of Two-Level Systems (TLS) \cite{Muller19}. Dissipation from this process originates from the tunneling of charge carrier between two energy configurations, forming an electric dipole that couples to the resonator. The possible microscopic mechanisms responsible for such processes are numerous, but the general expression for the quality factor degradation remains the same.\\
 A more refined model taking into account temperature-dependent TLS-TLS interactions (changing the state of one TLS has an influence on the state of other TLSs) leads to the expression~\cite{Crowley23}
 \begin{equation}
 Q_{\text{TLS}}(n,T) = Q_{\text{TLS,0}} \frac{\sqrt{1 + \left( \frac{n^{\beta_2}}{D T^{\beta_1}} \right) \tanh \left( \frac{\hbar \omega}{2k_BT} \right)}}{\tanh \left( \frac{\hbar \omega}{2k_BT} \right)}
 \end{equation}
 which we used to fit our data shown in Extended data Fig.~4a and b. Here $Q_{\text{TLS,0}}$ relates to the dissipation due to TLSs at low power and $T=0$, and the three phenomenological parameters $D, \, \beta_1, \, \beta_2$ characterize TLS thermal distribution and saturation.\\
 A fit of the $Q_i(T)$ curves using Eq.~(\ref{eq:TLS_1}) therefore involves seven fitting parameters; $A_{\text{QP}}$, $\Delta$, $Q_{\text{TLS,0}}$, $D$, $\beta_1$, $\beta_2$ and $Q_\text{other}$. The $Q_i(T)$ curves shown in Extended data Fig.~4b for different powers are fitted together at once with the same seven parameters. Note that, while we characterized the DC properties of the films (e.g. $T_c$), this does not allow us to extract the superconducting gap $\Delta$ and reduce the number of fitting parameters: in the pseudogap state of strongly disordered superconductors $\Delta$ and $T_c$ are unrelated.\\

 This model qualitatively reproduces the observed increase of quality factor with intra-cavity microwave power (see Extended Data Fig.~4a) and with temperature (Extended data Fig.~4b). Interestingly the low-$T$ part of the $Q_i(T)$ data ($T < 100$ mK) displays an initial decrease before increasing at higher temperature. This effect can be understood as as reduction of TLS coherence time upon increase of temperature and is included in the model above.

 \section{Dissipation in a:InO resonators is not due to surface dielectric loss}

 The microscopic origin of TLSs in our resonators is not known. Studies show that dielectric materials located at the interfaces surrounding the resonators host such defects and contribute to microwave dissipation~\cite{Muller19}. These dielectric materials come in thin layers (typically a few nm-thick), and originate from native oxides at the substrate or metal surfaces, or from resist residues. Here we show that the anomalously large dissipation in indium oxide resonators cannot be explained by surface dielectric loss alone.\\

 The coupling of TLS grows with electric field $\mathbf{E}$. Therefore the effect of TLSs on resonator dissipation directly translates into the fraction of electric field energy in a given volume $V_i$ of lossy material. We define a surface participation ratio $p_i$ in the volume $V_i$ as
 \begin{equation}
 p_i = \int_{V_i} \frac{\varepsilon_i}{2} |\mathbf{E_i(r)}|^2 / E_{\text{tot}} \, d\mathbf{r}
 \end{equation}
 where $\varepsilon_i$ and $\mathbf{E_i(r)}$ are the dielectric constant and electric field in the lossy volume $V_i$ respectively, and $E_{\text{tot}}$ is the total electric field energy in the entire space.\\
 To estimate the surface participation ratio between conductor and substrate $p_{\text{MS}}$ we follow \cite{Wang2015} and consider a 3 nm-thick dielectric layer with dielectric constant $\varepsilon_{\text{SiO}_2} = 10$ below the resonator. Using the electromagnetic simulation software Ansys HFSS we compute the electric field energy stored in the volume $V_{\text{MS}}$ for a given resonator mode, allowing the estimation of $p_{\text{MS}}$.\\
 In the microstrip geometry (displayed in Fig.~1) $p_{\text{MS}} \approx 1.5 \, 10^{-3}$. This is consistent with the results of \cite{Amin22} who used almost identical sample geometry and packaging.\\

 \begin{figure}[ht!]
            \includegraphics[width= 1\linewidth]{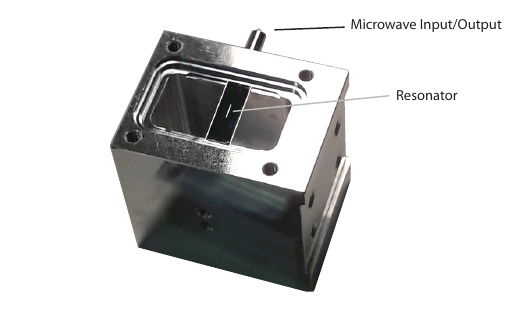}
\caption{\textbf{Aluminum 3D waveguide hosting an a:InO resonator.} The resonator TC040 (highlighted in white, in the center) is a 2 mm-long a:InO strip deposited on a silicon substrate. The waveguide has same dimensions as the ones studied in \cite{Grunhaupt18, Amin22}. It is closed by an aluminum lid (not shown) and sealed with an indium wire. Microwave signal enters and exits the waveguide through a coaxial cable. Input and output waves are separated using a microwave circulator. }
 \label{supp_mat_fig1}
 \end{figure}

 \subsection{Resonators in 3D waveguide}
 In order to drastically reduce $p_{\text{MS}}$ we designed a three-dimensional aluminum waveguide (see Fig.~S3) to which we couple an indium oxide resonator. For a $2~\text{mm}$-long, $1~\mathrm{\mu m}$-wide resonator with kinetic inductance $L_K = 1.32~\mathrm{nH / \square}$ on a silicon substrate electromagnetic simulations lead $p_{\text{MS}} \approx 3.17 \, 10^{-4}$, an order of magnitude smaller than the microstrip geometry.\\


Extended data Fig.~4 shows the relation between low-power quality factors and Metal-Substrate surface participation ratio for several materials (Al, Ta, TiN, grAl, a:InO and Al transmon qubits). If dissipation was mainly due to dielectric surface loss, quality factors should scale with participation ratio as $Q_i = \left[p_{\text{MS}} \tan \delta \right]^{-1}$. This expected scaling is represented in Extended data Fig.~4 with dashed lines for three values of the dielectric loss tangent $\tan \delta$.\\
While clean superconductors (Ta, Al) seem to follow the expected increase of $Q_i$ upon decrease of $p_{\text{MS}}$, we find that a reduction of $p_{\text{MS}}$ by one order of magnitude has virtually no effect on $Q_i$ for disordered superconductors such as grAl, TiN and a:InO. Strikingly, indium oxide resonators have a quality factor of $\sim 10^4$, orders of magnitude below the upper-bound set by the dielectric loss model, and irrespective of the sample holder geometry. This observation clearly points towards intrinsic bulk loss mechanisms for disordered superconducting resonators.
\\

\subsection{Effect of other interfaces and aluminum-capped resonators}
The discussion above focuses on the Metal-Substrate interface, while other regions of the system might also host dielectric layers. The interfaces between air and substrate (participation $p_{\text{AS}}$) or air and metal ($p_{\text{AM}}$) also play a role. While $p_{\text{AS}}$ is known to be of the order of magnitude as $p_{\text{MS}}$, and therefore cannot explain the unexpectedly large dissipation in a:InO, one must check the effects of oxides growing at the metal-air interface.\\

Microwave resonators made out of pure indium show poor performances $Q_i \sim 10^4$ as shown in \cite{McRae18}, with a possible interpretation being the growth of lossy indium oxide at the Metal-Air interface. To exclude such eventuality we fabricated an a:InO resonator (DP-res8) capped in-situ with a $\sim 1$ nm-thin aluminum layer right after film deposition. The capping layer acts as a protective covering for the resonator, where In$_2$O$_3$ native oxide cannot grow once in contact with air. Aluminum oxides have much lower loss, as seen in Extended data Fig.~4.\\
This device showed no improvement of quality factor, and displays the usual TLS-like behavior shown in Extended data Fig.~4, therefore ruling out the hypothesis of strongly dissipative metal-air interface.

\end{document}